\documentclass[nohyperref]{article}

\usepackage{microtype}
\usepackage{graphicx}
\usepackage{subfigure}
\usepackage{booktabs} % for professional tables
\usepackage{wrapfig}
\usepackage{multirow} 
\usepackage{multicol} 
\usepackage{hyperref}

\usepackage[accepted]{icml2023}

\usepackage{amsmath}
\usepackage{amssymb}
\usepackage{mathtools}
\usepackage{amsthm}
\usepackage{enumitem}
\usepackage{ulem}

\usepackage[capitalize,noabbrev]{cleveref}

\theoremstyle{plain}

\theoremstyle{definition}

\theoremstyle{remark}

\usepackage[textsize=tiny]{todonotes}

\icmltitlerunning{ICML2023}

\begin{document}

\twocolumn[
\icmltitle{Bidirectional Learning for Offline Model-based Biological Sequence Design}




\begin{icmlauthorlist}
\icmlauthor{Can (Sam) Chen}{mcgill,mila}
\icmlauthor{Yingxue Zhang}{huawei}
\icmlauthor{Xue Liu}{mcgill}
\icmlauthor{Mark Coates}{mcgill}
% \icmlauthor{Firstname2 Lastname2}{equal,yyy,comp}
% \icmlauthor{Firstname3 Lastname3}{comp}
\end{icmlauthorlist}

\icmlaffiliation{mila}{Mila - Quebec AI Institute}
\icmlaffiliation{mcgill}{McGill University}
\icmlaffiliation{huawei}{Huawei Noah's Ark Lab}

\icmlcorrespondingauthor{Can (Sam) Chen}{can.chen@mila.quebec}

\icmlkeywords{Machine Learning, ICML}

\vskip 0.3in
]


\printAffiliationsAndNotice{}  % leave blank if no need to mention equal contribution
%\printAffiliationsAndNotice{\icmlEqualContribution} 
% otherwise use the standard text.

\begin{abstract}
Offline model-based optimization aims to maximize a black-box objective function with a static dataset of designs and their scores.
In this paper, we focus on biological sequence design to maximize some sequence score.
A recent approach employs bidirectional learning, combining
a forward mapping for exploitation and a backward mapping for constraint, and it relies on the neural tangent kernel (NTK) of an infinitely wide network to build a proxy model.
Though effective, the NTK cannot learn features because of its
parametrization, and {its use prevents the incorporation of powerful pre-trained Language Models~(LMs) that can capture the rich biophysical information in millions of biological sequences}.
We adopt an alternative proxy model, adding a linear head to a
pre-trained LM, and propose a linearization scheme.
This yields a closed-form loss and also takes into account the biophysical information in the pre-trained LM.
In addition, the forward mapping and the backward mapping play different roles and thus deserve different weights during sequence optimization.
To achieve this, we train an auxiliary model and leverage its weak
supervision signal via a bi-level optimization framework to effectively learn how to balance the two mappings.
Further, by extending the framework, we develop the first 
learning rate adaptation module \textit{Adaptive}-$\eta$,  
which is compatible with all gradient-based algorithms for offline model-based optimization.
Experimental results on DNA/protein sequence design tasks verify the
effectiveness of our algorithm.
Our code is available~\href{https://anonymous.4open.science/r/BIB-ICML2023-Submission/README.md}{here.}

\end{abstract}

\section{Introduction}
Offline model-based optimization aims to maximize a black-box objective function with a static dataset of designs and their scores.
This offline setting is realistic since in many real-world scenarios we do not have interactive access to the ground-truth evaluation.
The design tasks of interest include material, aircraft, and biological sequence~\citep{trabucco2021conservative}. 
In this paper, we focus on biological sequence design, including DNA/protein sequence, with the goal of maximizing some specified property of these sequences.

A wide variety of methods have been proposed for biological sequence design, including
evolutionary algorithms~\citep{sinai2020adalead, ren2022proximal}, reinforcement learning methods~\citep{angermueller2019model}, Bayesian optimization~\citep{terayama2021black}, search/sampling using generative models~\citep{brookes2019conditioning, chan2021deep}, and GFlowNets~\citep{jain2022biological}.
Recently, gradient-based techniques have emerged as an effective alternative~\citep{trabucco2021conservative}.
These approaches first train a deep neural network~(DNN) on the static dataset as a proxy %to predict the property 
and then obtain the new designs by directly performing gradient ascent steps on the existing designs.
Such methods have been widely used in biological sequence design~\citep{norn2021protein,tischer2020design,linder2020fast}.
One obstacle is the out-of-distribution issue, where the trained proxy model is inaccurate for the newly generated sequences.

To mitigate the out-of-distribution issue, recent work proposes regularization of the model~\citep{trabucco2021conservative, yu2021roma, fu2021offline} or the design itself~\citep{can2022bidirectional}.
The first category focuses on training a better proxy by introducing inductive biases such as robustness~\citep{yu2021roma}.
The second category 
introduces bidirectional learning~\citep{can2022bidirectional}, which consists of a forward mapping and a backward mapping, to optimize the design directly.
Specifically, the backward mapping leverages the high-scoring design
to predict the static dataset and vice versa for the forward mapping,
which distills the information of the static dataset into the
high-scoring design. This approach achieves state-of-the-art
performances on a variety of tasks.
Though effective, the proposed bidirectional learning relies on the
neural tangent kernel~(NTK) of an infinite-width model to yield a
closed-form loss, which is a key component of its successful operation.
The NTK cannot learn features due to its parameterization~\citep{yang2021tensor} and thus the bidirectional learning cannot incorporate the wealth of biophysical information %that can be derived from
from Language Models~(LMs) pre-trained over a vast corpus of unlabelled sequences~\citep{elnaggar2020prottrans, ji2021dnabert}.

To solve this issue, we construct a proxy model by combining a finite-width pre-trained LM with an additional layer. We then linearize the resultant proxy model, inspired by the recent progress in deep linearization~\citep{achille2021lqf, dukler2021diva}.
This scheme not only yields a closed-form loss but also exploits the rich biophysical information that has been distilled in the pre-trained LM.
In addition, the forward mapping encourages exploitation in the sequence space and the backward mapping serves as a constraint to mitigate the out-of-distribution issue.
It is vital to maintain an appropriate balance between
exploitation and constraint, and this can vary across design
tasks as well as during the optimization process. We introduce a hyperparameter $\gamma$ to control the balance. However, how to properly select $\gamma$ is challenging because of the problem's offline optimization nature. Thus, we develop a bi-level optimization framework \textit{Adaptive}-$\gamma$.
In this framework, we train an auxiliary model and leverage its weak supervision signal to effectively update $\gamma$.
To sum up, we propose \textit{\textbf{BI}directional learning for  model-based \textbf{B}iological sequence design}~(\textbf{BIB}).

Since the offline nature prohibits standard cross-validation strategies for hyperparameter tuning, 
all current gradient-based offline model-based algorithms preset the learning
rate $\eta$. 
There is a danger of poor selection, and to address this, we propose an \textit{Adaptive}-${\eta}$ module,
which effectively adapts the learning rate $\eta$
via the weak supervision signal from the trained auxiliary model.
To the best of our knowledge, \textit{Adaptive}-${\eta}$ is the first learning rate adaptation module for gradient-based algorithms for offline model-based optimization. 
\textcolor{black}{We discuss the relationship between the \textit{Adaptive} modules and previous hyperparameter optimization work in Sec.~\ref{sec: related}.}
Experiments on DNA and protein sequence design tasks verify the effectiveness of {BIB} and \textit{Adaptive}-$\eta$.
%\Sam{should we add textit to BIB since we add it to adaptive eta}

To summarize, our contributions are three-fold:
\vspace{-8pt}
\begin{itemize}[leftmargin=*]
    \item Instead of adopting the NTK, we
    construct a proxy model by combining a pre-trained biological LM with an additional trainable layer. We then linearize the proxy model, leveraging the recent progress on deep linearization. This yields a closed-form loss computation in bidirectional learning and allows us to exploit the rich biophysical information distilled into the LM via pre-training over millions of biological sequences.
    \item We propose a bi-level optimization framework \textit{Adaptive}-$\gamma$ where we
    leverage weak signals from an auxiliary model to achieve a satisfactory trade-off between \textcolor{black}{design} exploitation and constraint. 
    \item
    We further extend this bi-level optimization framework to \textit{Adaptive}-$\eta$.
    As the first learning rate tuning scheme in offline model-based optimization, \textit{Adaptive}-$\eta$ allows learning rate adaptation for any gradient-based algorithm.
\end{itemize}

\section{Preliminaries}
\label{sec: pre}
\subsection{Offline Model-based Optimization}
Offline model-based optimization aims to find a design $\boldsymbol{X}$ to maximize some unknown objective $f(\boldsymbol{X})$.
This can be formally written as,
\begin{equation}
    \boldsymbol{X}^*=\arg\max_{\boldsymbol{X}}f(\boldsymbol{X})\,,
\end{equation}
where we have access to a size-${N}$ dataset  $\mathcal{D} = \{(\boldsymbol{X}_1, y_1)\},\cdots, \{(\boldsymbol{X}_N, y_{{N}})\}$ with $\boldsymbol{X}_i$ representing a certain design and $y_i$ denoting the design score.
In this paper,  $\boldsymbol{X}_i$ represents a biological sequence design, including DNA and protein sequences, and ${y_i}$ represents a property of the biological sequence such as the fluorescence level of the green fluorescent protein~\citep{sarkisyan2016local}.

\subsection{Biological Sequence Representation}
Following~\citep{norn2021protein, killoran2017generating, linder2021fast}, we adopt the position-specific scoring matrix to represent a length-$L$ protein sequence as $\boldsymbol{X} \in \mathbb{R}^{L \times 20}$, where $20$ represents $20$ different kinds of amino acids.
For a real-world protein, $\boldsymbol{X[l, :]}$~($0\le l \le L-1$) is a one-hot vector denoting one kind of amino acid.
During optimization, $\boldsymbol{X[l, :]}$ is a continuous vector and ${softmax}(\boldsymbol{X[l, :]})$ represents the probability distribution of all $20$ amino acids in the position $l$.
Similarly, for a DNA sequence, we have $\boldsymbol{X} \in \mathbb{R}^{L \times 4}$ where $4$ represents $4$ different DNA bases.

The protein sequence $\boldsymbol{X}$ is fed into the embedding layer of the LM, which produces the embedding,
\begin{equation}
    \boldsymbol{e} = EMB(softmax(\boldsymbol{X}))\,.
\end{equation}
The main block of the LM takes $\boldsymbol{e}$ as input and outputs biophysical features.
%for downstream tasks. 
%
The DNA LM, which adopts the $k$-mer representation, is a little different from protein LMs.
See Appendix~\ref{appendix: dna} for details.

\subsection{Gradient Ascent on Sequence}
A common approach to the posed offline model-based optimization is
to train a proxy $f_{\boldsymbol{\theta}}(\boldsymbol{X})$ on the offline dataset,
\begin{equation}
    \boldsymbol{\theta}^{*} = \arg \min_{\boldsymbol{\theta}}\frac{1}{{N}} \sum_{{i}=1}^{{N}}(f_{\boldsymbol{\theta}}(\boldsymbol{X}_i) - y_i)^2\,.
\end{equation}
Then we can obtain the high-scoring design $\boldsymbol{X}_h$ by ${T}$ gradient ascent steps:
\begin{equation}
    \boldsymbol{X}_{t+1} = \boldsymbol{X}_{t} + \eta \nabla_{\boldsymbol{X}} f_{\boldsymbol{\theta}^{*}}(\boldsymbol{X})|_{\boldsymbol{X} = \boldsymbol{X}_t}\,,  \quad \mathrm{for}\,\, t \in [0, {T}-1]\,,
    \label{eq: grad_ascent}
\end{equation}
where the high-scoring design $\boldsymbol{X}_{h}$ can be obtained as $\boldsymbol{X}_{{T}}$.

Considering the discrete nature of biological sequences, the input of $f_{\boldsymbol{\theta}}(\cdot)$ should be discrete one-hot vectors.
Following~\citep{norn2021protein}, we can perform the following conversion and
predict the score via:
\begin{align}
  \hat{\boldsymbol{X}_i} &= {softmax}(\boldsymbol{X}_i)\,,\\
  \boldsymbol{Z}_i &= onehot(argmax(\hat{\boldsymbol{X}_i}))\,,\\
  \hat{y} &= f_{\boldsymbol{\theta}}(\boldsymbol{Z}_i)\,.
\end{align}
Then the gradient regarding $\boldsymbol{X}_i$ can be approximated as,
\begin{equation}
    \frac{d f_{\boldsymbol{\theta}}(\boldsymbol{Z}_i)}{ d \boldsymbol{x}_i} \approx  \frac{d f_{\boldsymbol{\theta}}(\boldsymbol{Z}_i)}{ d \boldsymbol{z}_i} \frac{d \hat{\boldsymbol{x}}_i}{ d \boldsymbol{x}_i}\,,
\end{equation}
where we unroll the matrices $\boldsymbol{X}_i$, $\hat{\boldsymbol{X}}_i$ and $\boldsymbol{Z}_i$ as vectors $\boldsymbol{x}_i$,
$\hat{\boldsymbol{x}}_i$ and
$\boldsymbol{z}_i$ for notational convenience.
This approximation allows us to use backpropagation directly from the proxy to the sequence design $\boldsymbol{X}_i$. 
For brevity, we will still use $f_{\boldsymbol{\theta}}(\boldsymbol{X_i})$ to represent the proxy.

{\subsection{Bidirectional Learning}}

\begin{figure*}
\centering
    \includegraphics[width=1.6\columnwidth]{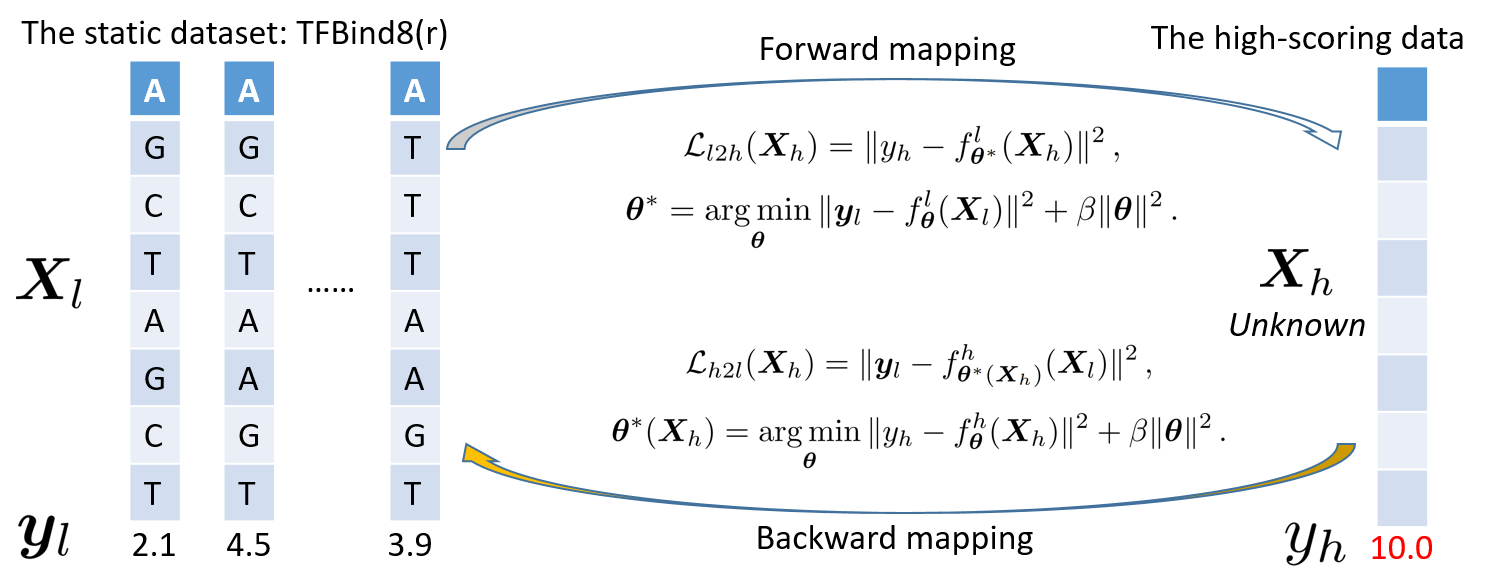}
\caption{{Illustration of bidirectional learning~\citet{can2022bidirectional} where ($\boldsymbol{X_l}$, $\boldsymbol{y_l}$) denotes the static dataset, $y_h$ is a large predefined target score and $\boldsymbol{X}_h$ is the high-scoring design we aim to find.
%predefine the target score $y_h$ as a large constant $10$.
%
% \citet{can2022bidirectional} predefine the target score $y_h$ as $10$ for all tasks and optimize the high-scoring design $\boldsymbol{X}_h$ by minimizing the bidirectional learning loss $\mathcal{L}(\boldsymbol{X}_h)= \mathcal{L}_{l2h}(\boldsymbol{X}_h) + \mathcal{L}_{h2l}(\boldsymbol{X}_h)$.
}
}
\label{fig: bdi}
%\vspace{-10pt}
\end{figure*}

{As shown in Figure~\ref{fig: bdi}, bidirectional learning~\citep{can2022bidirectional}, consists of two mappings: the forward mapping leverages the static dataset ($\boldsymbol{X_l}$, $\boldsymbol{y_l}$) to predict the score $y_h$ of the high-scoring design $\boldsymbol{X}_h$, and the backward mapping leverages the high-scoring design data ($\boldsymbol{X}_h$, $y_h$) to predict the static dataset ($\boldsymbol{X_l}$, $\boldsymbol{y_l}$).}
{The forward mapping loss is
\begin{equation}
     \mathcal{L}_{l2h}(\boldsymbol{X}_h) = \|y_h - {f}^l_{\boldsymbol{\theta}^*}(\boldsymbol{X}_h)\|^2\,,
\end{equation}
where $\boldsymbol{\theta}^*$ is given by 
\begin{equation}
    \label{eq: train_low}
    \boldsymbol{\theta}^* = \mathop{\arg\min}_{\boldsymbol{\theta}} \| \boldsymbol{y}_l - {f}^l_{\boldsymbol{\theta}}(\boldsymbol{X}_l)\|^2 + \beta \|\boldsymbol{\theta}\|^2\,,
\end{equation}
where $\beta>0$ is a regularization parameter. The backward mapping loss can be written as
\begin{equation}
    \label{eq: h2l_r}
    \mathcal{L}_{h2l}(\boldsymbol{X}_h) = \|\boldsymbol{y}_l - {f}^h_{\boldsymbol{\theta}^*(\boldsymbol{X}_h)}(\boldsymbol{X}_l)\|^2 \,,
\end{equation}
where $\boldsymbol{\theta}^*(\boldsymbol{X}_h)$ is given by 
\begin{equation}
    \boldsymbol{\theta}^*(\boldsymbol{X}_h) = \mathop{\arg\min}_{\boldsymbol{\theta}} \|y_h - {f}_{\boldsymbol{\theta}}^{h}(\boldsymbol{X}_h)\|^2 + \beta \|\boldsymbol{\theta}\|^2\,.
    \label{eq: inner}
\end{equation}
The high-scoring design $\boldsymbol{X}_h$ is optimized against the bidirectional learning loss $\mathcal{L}(\boldsymbol{X}_h)= \mathcal{L}_{l2h}(\boldsymbol{X}_h) + \mathcal{L}_{h2l}(\boldsymbol{X}_h)$.}

\section{Method}
In this section, we first illustrate how to leverage deep linearization to compute the bidirectional learning loss in a closed form.
Subsequently, we introduce a hyperparameter $\gamma$ to control the balance between the forward mapping and the backward mapping.
We then develop a novel bi-level optimization framework \textit{Adaptive}-$\gamma$, which leverages a weak supervision signal from an auxiliary model to effectively update $\gamma$.
Last but not least, we extend this framework to \textit{Adaptive}-$\eta$, which enables us to adapt the learning rate $\eta$ for
all gradient-based offline model-based algorithms.
{We summarize our method in Algorithm \ref{alg: bib}.}

\subsection{Deep Linearization for Bidirectional Learning}
\label{subsec: deeplinearization}
%

% The forward mapping loss can be written as
% \begin{equation}
%      \mathcal{L}_{l2h}(\boldsymbol{X}_h) = \|y_h - {f}^l_{\boldsymbol{\theta}^*}(\boldsymbol{X}_h)\|^2\,,
% \end{equation}
% \begin{equation}
%     \label{eq: train_low}
%     \boldsymbol{\theta}^* = \mathop{\arg\min}_{\boldsymbol{\theta}} \| \boldsymbol{y}_l - {f}^l_{\boldsymbol{\theta}}(\boldsymbol{X}_l)\|^2 + \beta \|\boldsymbol{\theta}\|^2\,.
% \end{equation}

%
% The backward mapping loss can be written as
% \begin{equation}
%     \label{eq: h2l_r}
%     \mathcal{L}_{h2l}(\boldsymbol{X}_h) = \|\boldsymbol{y}_l - {f}^h_{\boldsymbol{\theta}^*(\boldsymbol{X}_h)}(\boldsymbol{X}_l)\|^2 \,,
% \end{equation}
% \begin{equation}
%     \boldsymbol{\theta}^*(\boldsymbol{X}_h) = \mathop{\arg\min}_{\boldsymbol{\theta}} \|y_h - {f}_{\boldsymbol{\theta}}^{h}(\boldsymbol{X}_h)\|^2 + \beta \|\boldsymbol{\theta}\|^2\,.
%     \label{eq: inner}
% \end{equation}

In bidirectional learning, the backward mapping loss is intractable for a finite neural network, so~\citet{can2022bidirectional} employ a neural network with infinite width, which yields a closed-form loss via the NTK.
This however makes it impossible to incorporate the rich biophysical
information that has been distilled into a pre-trained LM~\citep{yang2021tensor}.
Considering this, we construct a proxy model by combining a finite-width pre-trained LM with an additional layer. 
We then linearize the resultant proxy model, inspired by the recent
progress in deep linearization which has established that an overparameterized DNN model is close to its linearization~\citep{achille2021lqf, dukler2021diva}.
Denote by $\boldsymbol{\theta_0}=(\boldsymbol{\theta_{pt}},\boldsymbol{\theta_{init}^{lin}}) \in\mathcal{R}^{D \times1}$ the proxy model parameters
derived by combining the parameters of the pre-trained LM $\boldsymbol{\theta_{pt}}$
and a random initialization of the linear layer
$\boldsymbol{\theta_{init}^{lin}}$. %where $\boldsymbol{\theta_0}\in\mathcal{R}^{D \times1}$. 
In this paper, we adopt the pre-trained DNABERT ~\citep{ji2021dnabert} and Prot-BERT~\citep{elnaggar2020prottrans} models, and compute the average of token embeddings as the extracted feature, which is fed into the linear layer to build the proxy.
\textcolor{black}{We also study how our method performs as a function of the pre-trained LM quality in Appendix~\ref{appendix: diff_prt_lm}.}
Then we can construct a linear approximation for the proxy model:
\begin{equation}
    f_{\boldsymbol{\theta}}(\boldsymbol{X}) \approx f_{\boldsymbol{\theta_0}}(\boldsymbol{X}) + \bigtriangledown_{\boldsymbol{\theta}}f_{\boldsymbol{\theta_0}}(\boldsymbol{X})\cdot(\boldsymbol{\theta} - \boldsymbol{\theta_0})\,,
\end{equation}
{where $f_{\boldsymbol{\theta}}(\boldsymbol{X}), f_{\boldsymbol{\theta_0}}(\boldsymbol{X}) \in\mathcal{R}$, $\bigtriangledown_{\boldsymbol{\theta}}f_{\boldsymbol{\theta_0}}(\boldsymbol{X}) \in\mathcal{R}^{1\times D}$ and $\bigtriangledown_{\boldsymbol{\theta}}f_{\boldsymbol{\theta_0}}(\boldsymbol{X}) \in\mathcal{R}^{D\times 1}$.}
Intuitively, if the fine-tuning does not significantly change $\boldsymbol{\theta_0}$, then this linearization is a good approximation.
%of the proxy model.
%
By leveraging this linearization, we can obtain a closed-form solution for Eq.(\ref{eq: inner}) as:
\begin{small}
\begin{equation}
    \begin{aligned}
        \boldsymbol{\theta}^*(\boldsymbol{X}_h) &= (\bigtriangledown_{\boldsymbol{\theta}}f_{\boldsymbol{\theta_0}}(\boldsymbol{X}_h)^\top \bigtriangledown_{\boldsymbol{\theta}}f_{\boldsymbol{\theta_0}}(\boldsymbol{X}_h) + \beta \boldsymbol{I})^{-1}\\ & \bigtriangledown_{\boldsymbol{\theta}}f_{\boldsymbol{\theta_0}}(\boldsymbol{X}_h)\top (y_h - f_{\boldsymbol{\theta_0}}(\boldsymbol{X}_h)) + \boldsymbol{\theta}_0.
    \end{aligned}
\end{equation}
\end{small}

Building on this result, we can compute the bidirectional learning loss as:
\begin{equation}
    \begin{aligned}
    \mathcal{L}_{bi}(\boldsymbol{X}_h)&=\frac{1}{2}(
    \|y_h-\boldsymbol{K}_{\boldsymbol{X}_h\boldsymbol{X}_l}(\boldsymbol{K}_{\boldsymbol{X}_l\boldsymbol{X}_l}+\beta \boldsymbol{I})^{-1}\\& (\boldsymbol{y}_l-f_{\boldsymbol{\theta_0}}(\boldsymbol{X}_l))\|^2
    + \|\boldsymbol{y}_l- \\&\boldsymbol{K}_{\boldsymbol{X}_l\boldsymbol{X}_h}(\boldsymbol{K}_{\boldsymbol{X}_h\boldsymbol{X}_h}+\beta \boldsymbol{I})^{-1}(y_h - f_{\boldsymbol{\theta_0}}(\boldsymbol{X}_h)) \|^2)
    \,,
    \label{eq: overall}
    \end{aligned}
\end{equation}
where $\boldsymbol{K}(\boldsymbol{X}_i, \boldsymbol{X}_j) = \bigtriangledown_{\boldsymbol{\theta}}f_{\boldsymbol{\theta_0}}(\boldsymbol{X}_i)\top \bigtriangledown_{\boldsymbol{\theta}}f_{\boldsymbol{\theta_0}}(\boldsymbol{X}_j).$
Following~\citep{dukler2021diva}, we can also only linearize the last layer of the network for simplicity, which defines the following kernel,
\begin{equation}
    \boldsymbol{K}(\boldsymbol{X}_i, \boldsymbol{X}_j) = {BERT} (\boldsymbol{X}_i)^\top {BERT} (\boldsymbol{X}_j),
\end{equation}
where ${BERT} (\boldsymbol{X})$ denotes the feature of the sequence $\boldsymbol{X}$ extracted by BERT.
Its kernel nature makes this approach suitable for small-data tasks~\citep{arora2019harnessing}, especially in drug discovery with high labeling cost of DNA/proteins.
%
%\textcolor{black}{Bidirectional learning is inherited from \cite{can2022bidirectional} and the novel point in this subsection is deep linearization.}

\subsection{Adaptive-\texorpdfstring{$\gamma$}{Lg}}
\begin{algorithm}[tb]
   \caption{Bidirectional Learning for Offline Model-based Biological Sequence Design}\label{alg: bib}
\textbf{Input:} The static dataset $\mathcal{D}$ = $(\boldsymbol{X}_l, \boldsymbol{y}_l)$, the predefined
target score $y_h = 10$, \# iterations $T$, the pre-trained biological LM parameterized by $\boldsymbol \theta_0$, the auxiliary model $f_{aux}(\cdot)$, the regularization $\beta$.%\\
%\textbf{Output:} $\boldsymbol{X}_h^{*}$ high-scoring design. 
% \begin{algorithmic}
%    %\STATE {\bfseries Input:} data $x_i$, size $m$
%    %\REPEAT
%    \STATE Initialize $\boldsymbol{X}_0$ as the sequence with the highest score in $\mathcal{D}$ 
%    \FOR{$i=1$ {\bfseries to} $m-1$}
%    \IF{$x_i > x_{i+1}$}
%    \STATE Swap $x_i$ and $x_{i+1}$
%    \ENDIF
%    \ENDFOR
%    %\UNTIL{$noChange$ is $true$}
% \end{algorithmic}
\begin{algorithmic}
\STATE Initialize $\boldsymbol{X}_0$ as the sequence with the highest score in $\mathcal{D}$ 
%\For{$\tau \leftarrow0$ \textbf{to} $T-1$}
\FOR{$\tau=0$ {\bfseries to} $T-1$}
\STATE Leverage Adaptive-$\gamma$ in Sec~\ref{subsec: gamma_learn} to update the balance $\gamma$ by Eq. (\ref{eq: gamma_learn})
\IF{Adapt learning rate} \STATE{Leverage Adaptive-$\eta$ in Sec~\ref{subsec: eta_learn} to update the learning rate $\eta$ by Eq. (\ref{eq: lr_learn})}
\ENDIF
\STATE Optimize $\boldsymbol{X}$ by minimizing the bidirectional learning loss $\mathcal{L}_{bi}(\boldsymbol{X}_\tau, \gamma)$ in Eq. ($\ref{eq: overall_}$):
\STATE \hspace{0.45cm} $\boldsymbol{X}_{\tau+1} = \boldsymbol{X}_{\tau} - \eta {OPT}(\nabla_{\boldsymbol{X}} \mathcal{L}_{bi}(\boldsymbol{X}_\tau, \gamma))$
\ENDFOR
\STATE Return $\boldsymbol{X}_h^{*} = \boldsymbol{X}_T$
\end{algorithmic}
\end{algorithm}
\label{subsec: gamma_learn}
The forward mapping and the backward mapping play different roles in
the sequence optimization process: the forward mapping encourages the
high-scoring sequence to search for a higher target score (exploitation) and the backward mapping serves as a constraint.
Since different sequences require different degrees of
constraint, we introduce an extra hyperparameter $\gamma \in [0, 1]$
to control the balance between the corresponding terms in the loss
function:
\begin{equation}
    \begin{aligned}
    \mathcal{L}_{bi}(\boldsymbol{X}_h, \gamma)&= \gamma \mathcal{L}_{l2h}(\boldsymbol{X}_h) + (1-\gamma) \mathcal{L}_{h2l}(\boldsymbol{X}_h)
    \,.
    \label{eq: overall_}
    \end{aligned}
\end{equation}
Thus $\gamma = 1.0$ corresponds to the forward mapping alone,
$\gamma = 0$ results in backward mapping, and $\gamma=0.5$ leads to
the bidirectional loss of~\citep{can2022bidirectional}.

It is non-trivial to determine the most suitable value for $\gamma$
since we do not know the ground-truth score for a new design. 
One possible solution is to train an auxiliary $f_{aux}(\cdot)$ 
%on the offline dataset 
to serve as a proxy evaluation.
A reasonable auxiliary is a simple regression model fitted to the offline dataset.
Although this auxiliary model cannot yield ground-truth scores, it can
provide weak supervision signals to update $\gamma$, since the auxiliary model and the bidirectional learning provide complementary information.
This is similar to co-teaching~\citep{han2018co} where two models leverage each other's view.

Formally, we introduce the \textit{Adaptive}-$\gamma$ framework.
Given a good choice of $\gamma$, the produced $\boldsymbol{X}_h$ is
expected to have a high score $f_{aux}(\boldsymbol{X}_h)$, based on which we can choose $\gamma$.
To make the search for $\gamma$ more efficient, we can formulate
this process as a bi-level optimization problem:
\begin{alignat}{2}
\gamma^* &= \mathop{\arg\max}_{\gamma}  f_{aux}(\boldsymbol{X}_h^{*}(\gamma))\,,\\
\mbox{s.t.}\quad  \boldsymbol{X}_h^{*}(\gamma) &= \mathop{\arg\min}_{\boldsymbol{X}_h} \mathcal{L}_{bi}(\boldsymbol{X}_h, \gamma)
    \,.
\end{alignat}
We can then use the hyper-gradient $\frac{\partial f_{aux}(\boldsymbol{X}_h^{*}(\gamma))}{\partial \gamma}$ to update $\gamma$.
Specifically, the inner level solution can be approximated via a gradient descent step with a learning rate $\eta$:
\begin{equation}
    \boldsymbol{X}_h^{*}(\gamma) = \boldsymbol{X}_h - \eta \frac{d \mathcal{L}_{bi}(\boldsymbol{X}_h, \gamma)}{d \boldsymbol{X}_h^\top}.
\end{equation}
For the outer level, we update $\gamma$ by hyper-gradient ascent:
\begin{equation}
\begin{aligned}
    \label{eq: gamma_learn}
    \gamma &= \gamma + \eta^{'}\frac{d f_{aux}(\boldsymbol{X}_h^{*}(\gamma))}{ d \gamma} = \gamma + \eta^{'}\frac{d f_{aux}(\boldsymbol{X}_h)}{d \boldsymbol{x}_h} \frac{d \boldsymbol{x}_h^{*}(\gamma)}{d \gamma} \\
    &= \gamma + \eta^{'}\eta\frac{d f_{aux}(\boldsymbol{X}_h)}{d \boldsymbol{x}_h}\frac{d \mathcal{L}_{h2l}(\boldsymbol{X}_h) - \mathcal{L}_{l2h}(\boldsymbol{X}_h)}{d \boldsymbol{x}_h^\top}\,,  
\end{aligned}
\end{equation}
where we unroll the matrix form $\boldsymbol{X}_h$ as a vector form $\boldsymbol{x}_h$ for better illustration.

\subsection{Adaptive-\texorpdfstring{$\eta$}{Lg}}
\label{subsec: eta_learn}
We now extend the \textit{Adaptive}-$\gamma$ framework to \textit{Adaptive}-$\eta$.
As the first learning rate adaptation module for offline model-based optimization, \textit{Adaptive}-$\eta$ is compatible with all gradient-based algorithms and can effectively finetune the learning rate $\eta$ via the auxiliary model's weak supervision signal.
All gradient-based methods that maximize $\mathcal{L}_{\boldsymbol{\theta}}(\boldsymbol{X})$ with respect to $\boldsymbol{X}$ have the following general form:
\begin{small}
\begin{equation}
    \boldsymbol{X}_{t+1} = \boldsymbol{X}_{t} + \eta {OPT}(\nabla_{\boldsymbol{X}} \mathcal{L}_{\boldsymbol{\theta}}(\boldsymbol{X})|_{\boldsymbol{X} = \boldsymbol{X}_t})\,,  \quad \mathrm{for}\,\, t \in [0, \rm{T}-1]\,,
    \label{eq: grad_ascent_general}
\end{equation}
\end{small}
where $\eta$ represents the learning rate of the optimizer.
For methods such as simple gradient ascent (Grad), COMs~\citep{trabucco2021conservative},
ROMA~\citep{yu2021roma} and NEMO~\citep{fu2021offline},
$\mathcal{L}_{\boldsymbol{\theta}}(\cdot)$ is related to the proxy model
$f_{\boldsymbol{\theta}}(\cdot)$; for BDI~\cite{can2022bidirectional}
and our proposed method, BIB, $\mathcal{L}_{\boldsymbol{\theta}}(\cdot)$
is the negative of the bidirectional learning loss, i.e., 
$\mathcal{L}_{\boldsymbol{\theta}}=-\mathcal{L}_{bi}$.

Though the learning rate $\eta$ can be adapted in some optimizers such as Adam~\citep{kingma2014adam}, these adaptations rely on only the past optimization history and do not consider the weak supervision signal from the auxiliary model. 
Our \textit{Adaptive}-$\eta$ optimizes $\eta$ by solving:
\begin{equation}
    \label{eq: lr_learn}
    \eta^* = \mathop{\arg\max}_{\eta}  f_{aux}(\boldsymbol{X}_h^{*}(\eta))\,,
\end{equation}
where $\eta$ can be updated via gradient ascent methods.
Considering the sequence optimization procedure is highly sensitive to the learning rate $\eta$, we reset $\eta$ to $\eta_0$ at each iteration and update $\eta$ from $\eta_0$, 
\begin{equation}
    \eta = \eta_0 - \eta^{'} \frac{d f_{aux}(\boldsymbol{X}_h^{*}(\eta))}{d \eta}.
\end{equation}
In general, this stabilizes the optimization procedure.

\section{Experiments}
We conduct extensive experiments on DNA and protein design tasks, and aim to answer three research questions: 
(1) How does BIB compare with state-of-the-art algorithms?
(2) Is every design component necessary in BIB?
(3) Does the \textit{Adaptive}-$\eta$ module improve gradient-based methods?

\subsection{Benchmark}
%W
\label{subsec: benchmark}
We conduct experiments on two DNA tasks: TFBind8(r) and TFBind10(r), following~\citep{can2022bidirectional} and three protein tasks: avGFP, AAV and E4B, in \citep{ren2022proximal} which have the most data points. See Appendix~\ref{appendix: dataset} for more details on task definitions and oracle evaluations.
\textcolor{black}{We also study how our method performs with different task dataset sizes in Appendix~\ref{appendix: diff_data_size}.}

Following~\citep{trabucco2021conservative}, we select the top ${N}=128$ most promising sequences for each comparison method.
Among these sequences, we report the maximum  normalized ground truth score as the evaluation metric following~\citep{ren2022proximal}.

\subsection{Comparison Methods}
We compare BIB with two groups of baselines: the gradient-based methods and the non-gradient-based methods.
For a fair comparison, the pre-trained LM is used for all methods involving a proxy~{and we don't finetune the LM.}
The gradient-based methods include: 
1) Grad: gradient ascent on existing sequences to obtain new sequences; 
2) COMs~\citep{trabucco2021conservative}: lower bounds the DNN model
by the ground-truth values and then applies gradient ascent;
3) ROMA~\citep{yu2021roma}: incorporates a smoothness prior into the DNN model before gradient ascent steps;
4) NEMO~\citep{fu2021offline}: leverages the normalized maximum-likelihood estimator to bound the distance between the DNN model and the ground-truth values;
5) BDI~\citep{can2022bidirectional}: adopts the infinitely wide NN and its NTK to compute bidirectional learning loss.

The non-gradient-based methods include:
1) BO-qEI~\citep{wilson2017reparameterization}: builds an acquisition function for sequence exploration;
2) CMA-ES~\citep{hansen2006cma}: estimates the covariance matrix to adjust the sequence distribution towards the high-scoring region;
3) AdaLead~\citep{sinai2020adalead}: performs a hill-climbing search on the proxy and then queries the sequences with high predictions;
4) CbAS~\citep{brookes2019conditioning}: builds a generative model for sequences above a property threshold and gradually adapts the distribution by increasing the threshold;
5) PEX~\citep{ren2022proximal}: prioritizes the evolutionary search for protein sequences with low mutation counts;
6) GENH~\citep{chan2021deep}: enhances the score through a learned latent space.

\subsection{Training Details}
\label{subsec: training_details}
We follow the training setting in~\citep{can2022bidirectional} if not specified.
We choose $OPT$ as the Adam optimizer~\citep{kingma2014adam} for all gradient-based methods.
We implement the auxiliary model as a linear layer with the feature from the pre-trained LM. %into a linear layer to output the score.
%
%We train five auxiliary models and use the emsemble for all methods for a fair comparison. 
%
We set the number of iterations $T$ as $25$ for all experiments following~\citep{norn2021protein} and $\eta_0$ as $0.1$ following~\citep{can2022bidirectional}.
%, we set the number of iterations $T$ as $25$ for all experiments.
%and set $\eta_0$ as $0.1$ following~\citep{can2022bidirectional}.
%and initialize the protein sequence as a gaussian vector.
%
%See Appendix~\ref{appendix: init} for details.
%
%We train the auxiliary model which feeds the feature from the pre-trained LM into a linear layer to output the score.
%
%
%We adopt the Adam optimizer~\citep{kingma2014adam} for all gradient-based methods.
%
We run every setting over $16$ trials and report the mean and standard deviation.
%
%We use Pytorch~\citep{paszke2019pytorch} to run all experiments on one V100 GPU.
%
See Appendix~\ref{appendix: init} for other training details.

%\vspace{-10pt}

\subsection{Results and Analysis}
\begin{figure}
    \centering
    \includegraphics[scale=0.18]{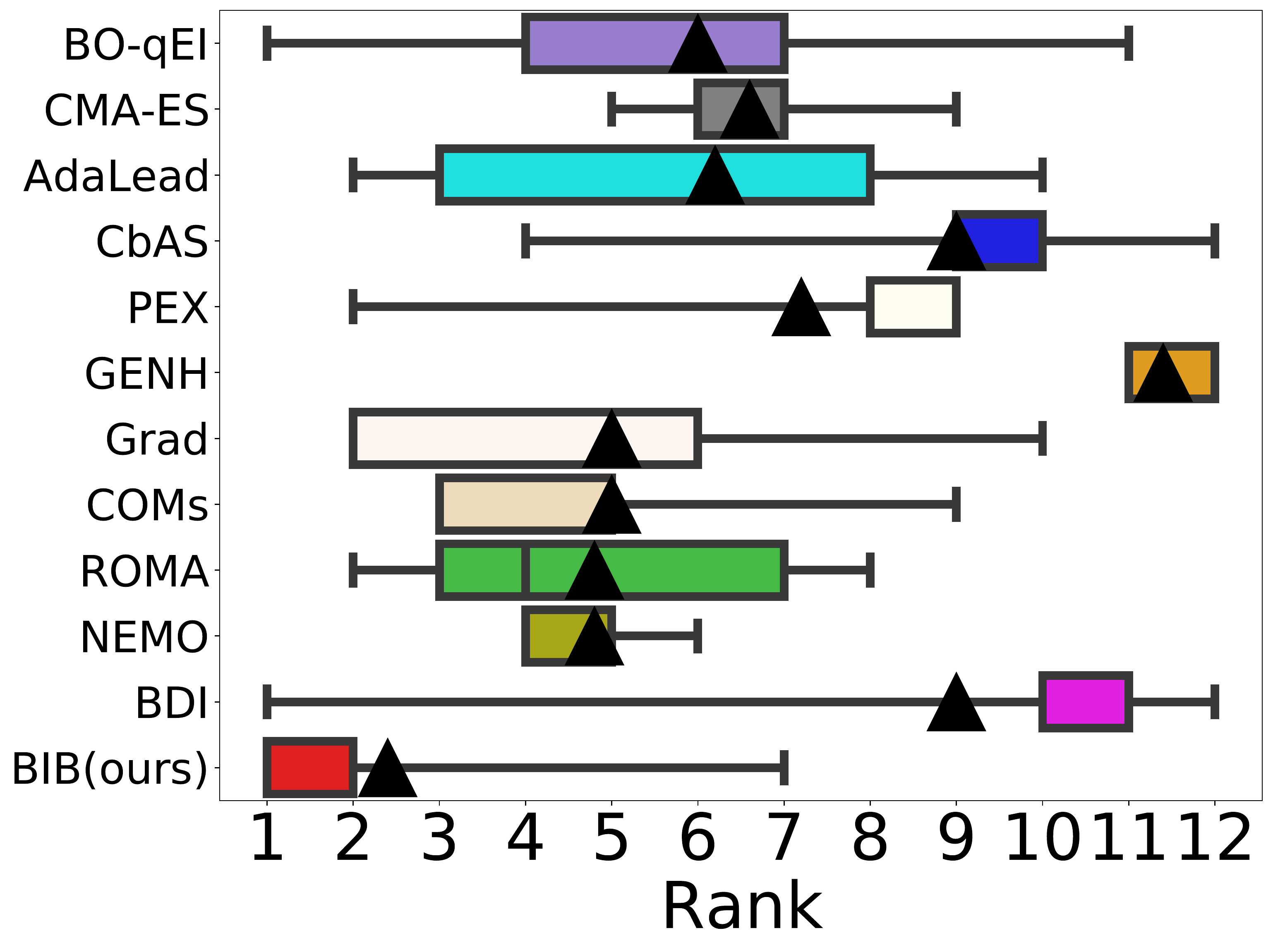}
    \vspace{-10pt}
    \caption{Rank minima and maxima are represented by whiskers; vertical lines and black triangles denote medians and means.}
    %\vspace{-10pt}
    \label{fig: ranking}
    \vspace{-20pt}
\end{figure}

We report all experimental results in Table~\ref{tab: continuous} and plot the ranking statistics in Figure~\ref{fig: ranking}.
We make the following observations.
(1) As shown in Table~\ref{tab: continuous}, BIB consistently
outperforms the Grad method on all tasks, which demonstrates that our BIB can effectively mitigate the out-of-distribution issue.
(2) Furthermore, BIB outperforms BDI on $4$ out of $5$ tasks, which demonstrates the effectiveness of the pre-trained biological LM over NTK. 
\textcolor{black}{We have conducted experiments, reported in Appendix~\ref{appendix: rank_perf}, to verify that the ranking of prediction performance is NN $>$ linearized pre-trained LM $>$ NTK.}
{The reason why BDI outperforms BIB on TFBind10(r) may be that short sequences do not rely much on the rich sequential information from the pre-trained LM.} 
(3) As shown in Figure~\ref{fig: ranking}, the gradient-based methods  generally perform better than the non-gradient-based methods, as also observed by~\cite{trabucco2021conservative}.
%\vspace{-20pt}
% \begin{wrapfigure}[14]{r}{0.35\textwidth}
%     \centering
%     %\vspace{-10pt}
%     \includegraphics[scale=0.14]{Figures/rank.pdf}
%     %\captionsetup{font={small}}
%     \vspace{-10pt}
%     \caption{Rank minima and maxima are represented by whiskers; vertical lines and black triangles denote medians and means.}
%     \label{fig: ranking}
% \end{wrapfigure}
(4) The gradient-based methods are inferior for the AAV task. One possible reason is that the design space of AAV ($20^{28}$) is much
smaller than those of avGFP ($20^{239}$) and E4B ($20^{102}$), which makes the generative modeling and evolutionary algorithms more suitable.
(5) This conjecture is also supported by the experimental results on two DNA design tasks.
%the advantage of \textit{gradient-based methods over non-gradient-based methods} is greater for TFBind10(r) than TFBind8(r).
%
{We compute the average ranking of gradient-based methods and non-grad-based methods on TFBind10(r) as $3.5$ and $9.5$, respectively, and the average ranking of gradient-based methods and non-grad-based methods on TFBind8(r) as $5.8$ and $6.8$, respectively. The advantage of gradient-based methods are larger ($9.5-3.5 = 6.0$) in TFBind10(r) than that ($6.8-5.8 = 1.0$) in TFBind8(r).}
(6) The generative modeling methods CbAS and GENH yield poor results
on all tasks, probably because the high-dimensional data distribution
is very hard to model.
(7) Overall, BIB attains the best performance in $3$ out of $5$ tasks and achieves the best ranking results as shown in Table~\ref{tab: continuous} and Figure~\ref{fig: ranking}.

\begin{figure*}
\centering
    \subfigure[TFBind8(r).]
    {
    \includegraphics[width=0.630\columnwidth]{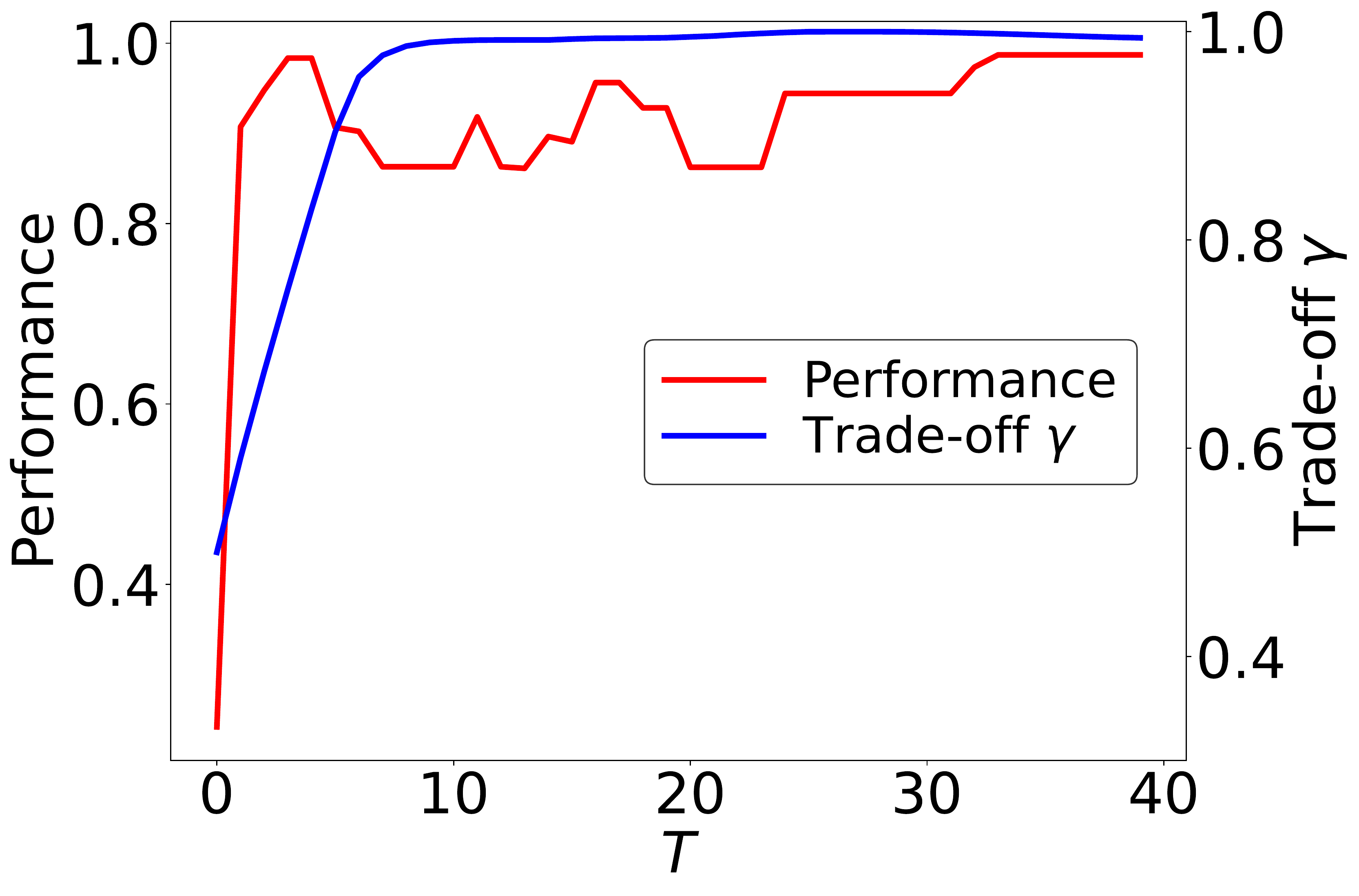}
    %\label{fig: TFBind8}
    }
    \subfigure[avGFP.]
    {
    \includegraphics[width=0.630\columnwidth]{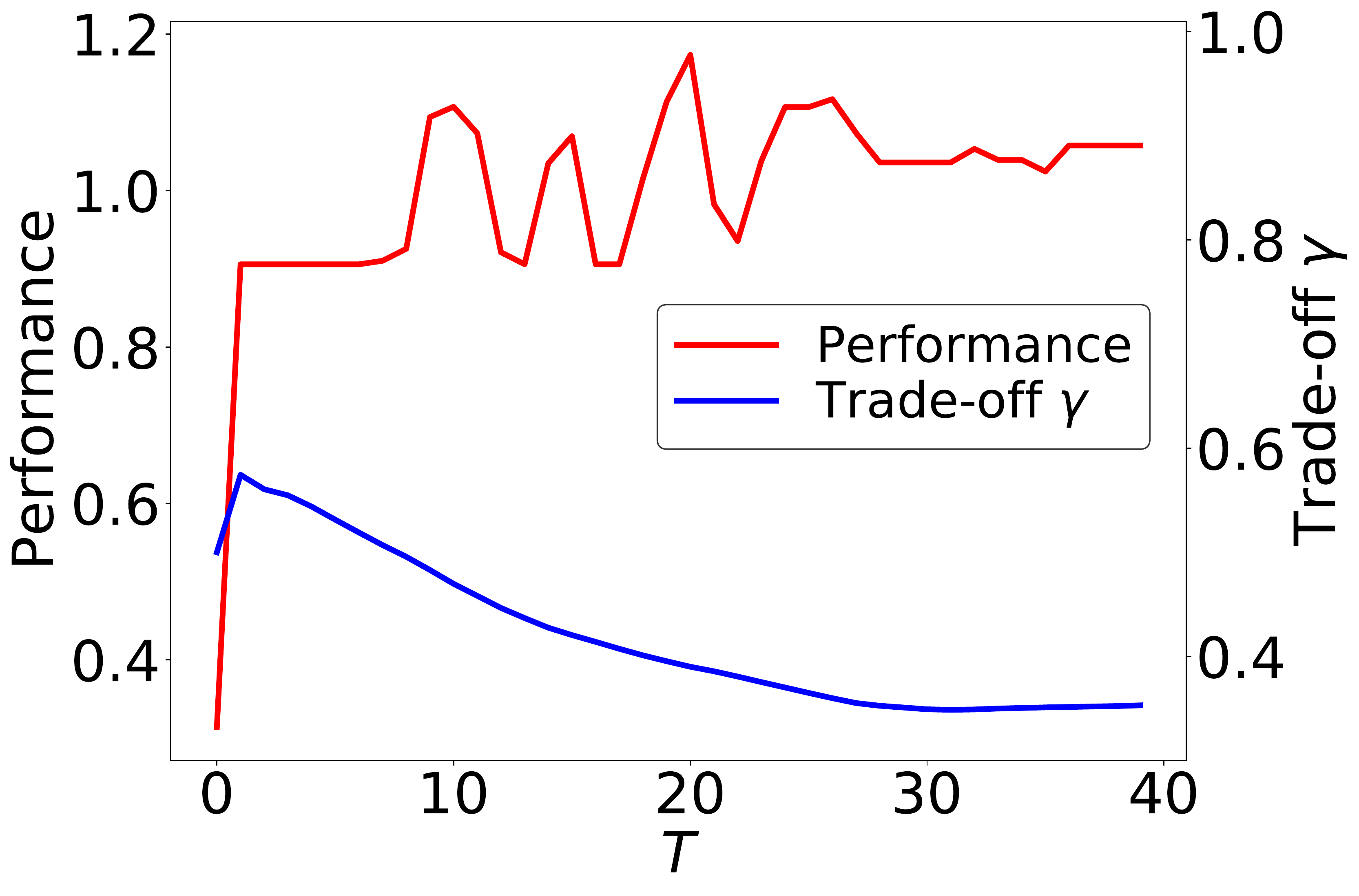}
    }
    \subfigure[E4B.]
    {
    \includegraphics[width=0.630\columnwidth]{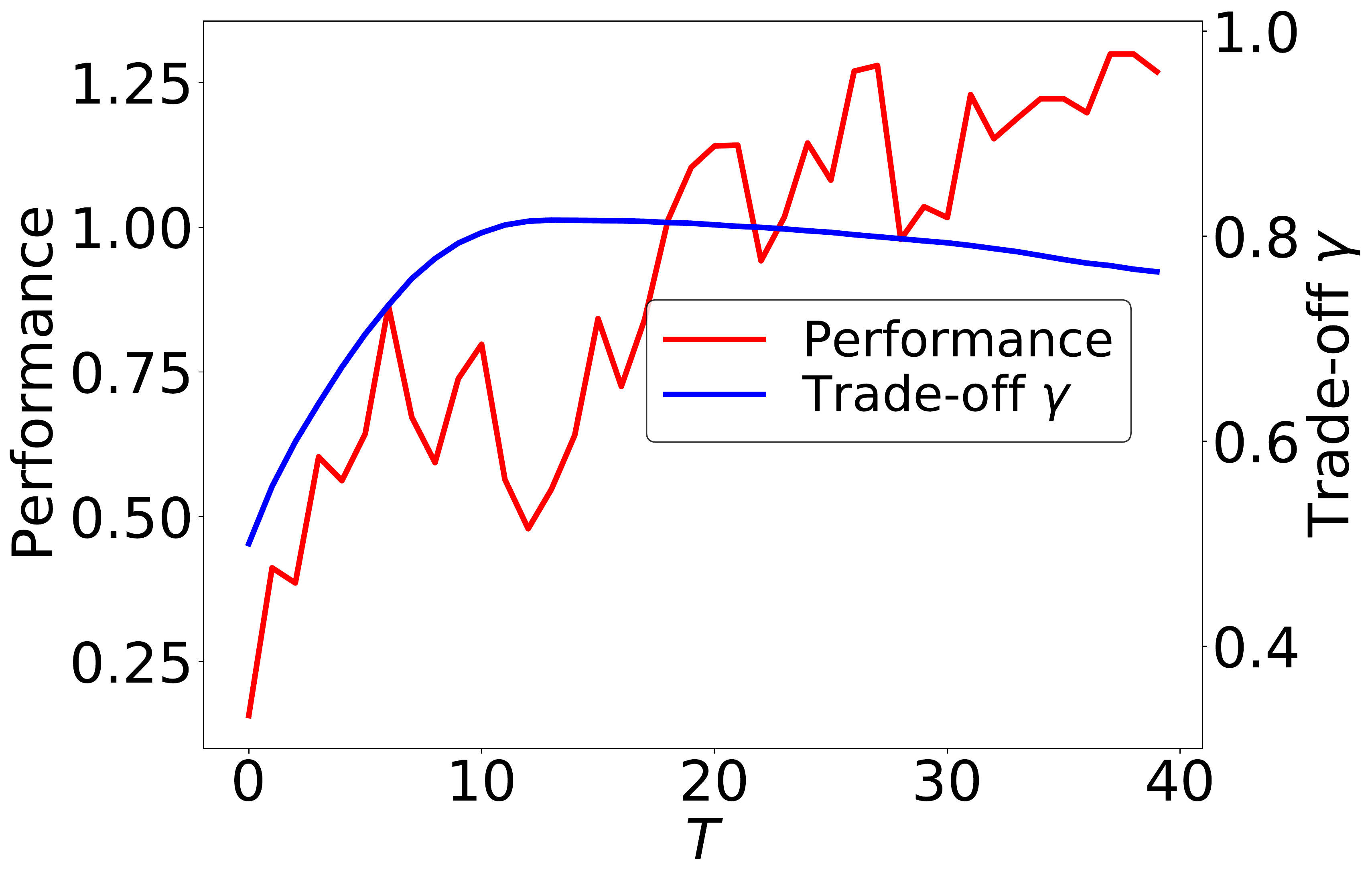}
    }
\vspace{-10pt}
\caption{Trend of performance and trade-off $\gamma$ as a function of $T$.}
\label{fig: trend}
\vspace{-10pt}
%\vspace{-20pt}
\end{figure*}

We also visualize the trend of performance~(the maximum  normalized ground truth score) and trade-off $\gamma$ as a function of $T$ on TFBind8(r) in Figure~\ref{fig: trend}(a) and avGFP in Figure~\ref{fig: trend}(b).
The performance generally increases with the time step $T$ and then stabilizes, which demonstrates the effectiveness and robustness of BIB.
Furthermore, we find that the $\gamma$ values of TFBind8(r) and avGFP
generally increase at first.
This means that BIB reduces the impact of the
constraint to encourage a more aggressive search for a high target value during the initial phase.
Then $\gamma$ of TFBind8(r) continues to increase while the $\gamma$ of avGFP decreases.
We conjecture that the difference is caused by the sequence length.
Small mutations of a biological sequence are enough to yield a good candidate~\citep{ren2022proximal}.
For the length-$239$ protein in avGFP, dramatic mutations 1) are not
necessary and 2) can easily lead to out-of-distribution points.
The weak supervision signal from the auxiliary model therefore
encourages a tighter constraint towards the static dataset.
By contrast, the DNA sequence is relatively short and a more widespread search of the sequence space can yield better results.
To investigate this conjecture, we further visualize the trend of E4B in Figure~\ref{fig: trend}(c).
E4B also has long sequences ($102$) and we can observe its similar
first-increase-then-decrease trend, although it is not as pronounced.

\begin{table*}
	\centering
	\caption{{Experimental results (maximum  normalized ground truth score) for comparison.}}  
	\label{tab: continuous}
	\scalebox{0.9}{
	\begin{tabular}{cccccccc}
		\specialrule{\cmidrulewidth}{0pt}{0pt}
		Method & TFBind8(r) & TFBind10(r) & avGFP & AAV & E4B & Rank Mean & Rank Median\\
		\specialrule{\cmidrulewidth}{0pt}{0pt}
		$\mathcal{D}$(\textbf{best}) & $0.242 $ & $0.248$ & $0.314$& $0.452$& $0.224$ &  &  \\
		BO-qEI & $0.940 \pm {0.032}$ & $0.595 \pm {0.028}$ & $0.888 \pm {0.015}$& $\textbf{0.591} \pm \textbf{0.002}$& $0.436 \pm {0.004}$ & $6.0/12$ & $7.0/12$ \\
		CMA-ES & $0.930\pm {0.034}$ & $0.617\pm {0.031}$ & $0.909 \pm {0.004}$& $0.470\pm {0.006}$& $0.748\pm {0.009}$ & $6.6/12$ & $6.0/12$ \\
		AdaLead & $\uline{0.941} \pm \uline{ {0.032}}$ & $0.602\pm {0.028}$ & $0.885 \pm {0.016}$& $0.581\pm {0.002}$& $0.433\pm {0.003}$ & $6.2/12$ & $8.0/12$ \\
		CbAS & $0.878\pm {0.049}$ & $0.610\pm {0.035}$ & $0.785 \pm {0.057}$& $0.543\pm {0.002}$& $0.349\pm {0.003}$ & $9.0/12$ & $10.0/12$ \\
		PEX & $0.924\pm {0.041}$ & $0.612\pm {0.026}$ & $0.874 \pm {0.033}$& $\uline{0.588} \pm \uline{0.002}$ & $0.397 \pm {0.004}$ & $7.2/12$ & $8.0/12$ \\
		GENH & $0.323\pm {0.000}$ & $0.448\pm {0.000}$ & $0.793\pm {0.000}$& $0.452\pm {0.000}$& $0.228\pm {0.000}$ & $11.4/12$ & $11.0/12$ \\
		\specialrule{\cmidrulewidth}{0pt}{0pt}
		Grad & $\uline{0.941} \pm \uline{0.026}$ & $0.630\pm {0.029}$ & $0.913 \pm {0.027}$& $0.463\pm {0.005}$& $\uline{1.219} \pm \uline{0.061}$ & $5.0/12$ & $5.0/12$ \\
		COMs & $0.921\pm {0.039}$ & {$0.637\pm {0.065}$} & $0.938\pm {0.048}$& $0.511\pm {0.005}$& $0.829\pm {0.026}$ & $5.0/12$ & $5.0/12$ \\
		ROMA & $0.926\pm {0.032}$ & $0.634\pm {0.061}$ & $\uline{0.975} \pm \uline{0.133}$& $0.471 \pm {0.005}$& $1.198 \pm {0.042}$ & $4.8/12$ & $4.0/12$ \\
		NEMO & $0.930\pm {0.038}$ & $0.632\pm {0.024}$ & $0.914 \pm {0.026}$& $0.505\pm {0.005}$& $1.036\pm {0.046}$ & $4.8/12$ & $5.0/12$ \\
		BDI & {$0.823\pm {0.000}$} & $\textbf{0.678}\pm \textbf{0.000}$ & {$0.873\pm {0.000}$}& $0.452\pm {0.000}$& $0.224\pm {0.000}$ & $9.0/12$ & $11.0/12$ \\
		\specialrule{\cmidrulewidth}{0pt}{0pt}
		\specialrule{\cmidrulewidth}{0pt}{0pt}
		\textbf{BIB}$_{\mathrm{(ours)}}$ & $\textbf{0.952} \pm \textbf{0.033}$ & $\uline{0.639} \pm \uline{0.032}$ & $\textbf{1.060} \pm \textbf{0.016}$& $0.501 \pm {0.007}$& $\textbf{1.255} \pm \textbf{0.029}$ & $2.4/12$ & $1.0/12$ \\	\specialrule{\cmidrulewidth}{0pt}{0pt}
	\end{tabular}
	}
\end{table*}

\begin{table}
	\centering
	\caption{Ablation studies on BIB components.}  
	\label{tab: ablation}
	\scalebox{0.65}{
	\begin{tabular}{cccccccc}
		\specialrule{\cmidrulewidth}{0pt}{0pt}
		\specialrule{\cmidrulewidth}{0pt}{0pt}
		Task &   $\gamma=0.0$ &   $\gamma=1.0$ & $\gamma=0.5$  & $\gamma=0.5$ + Joint  & BIB & BIB + Ada-$\eta$ \\
		\specialrule{\cmidrulewidth}{0pt}{0pt}
		TFB8(r) & ${0.936} $ & $0.933$ & ${0.947} $& $0.935$ &  $0.952$&  $0.956$\\
		TFB10(r) & ${0.611} $ & $0.637$ & ${0.616} $& $0.622$ &  $0.639$&  $0.639$\\
		\specialrule{\cmidrulewidth}{0pt}{0pt} 
		avGFP & ${0.920} $ & $0.966$ & ${1.018} $& $1.006$ &  $1.060$&  $1.082$\\
		AAV & ${0.449} $ & $0.458$ & ${0.480} $& $0.420$ &  $0.501$&  $0.525$\\
		E4B & $ {0.778} $ & $0.903$ & ${1.198} $& $1.176$ &  $1.255$&  $1.301$\\
		\specialrule{\cmidrulewidth}{0pt}{0pt}
		\specialrule{\cmidrulewidth}{0pt}{0pt}
	\end{tabular}
	}
\end{table}
\vspace{-10pt}

\subsection{Ablation Studies}
\label{subsec: abalation}
In this subsection, we conduct ablation studies to verify the effectiveness of the forward mapping, the backward mapping, and the \textit{Adaptive}-$\gamma$ module of BIB.
\textcolor{black}{We stress that forward mapping and backward mapping are not our contributions. In this paper, we propose deep linearization for bidirectional mappings. This section explores whether both mappings remain effective after deep linearization has been performed.}
We report the experimental results in Table~\ref{tab: ablation}.

\noindent \textbf{Forward mapping \& Backward mapping.}
We can observe that bidirectional learning~($\gamma=0.5$) performs better than both forward mapping~($\gamma=1.0$) and backward mapping~($\gamma=0.0$) alone in most tasks, which demonstrates the effectiveness of forward mapping and backward mapping.
The advantage of \textit{bidirectional mappings over the forward mapping} is larger in the long-sequence tasks like avGFP~(238) and E4B~(102) compared with the short-sequence tasks.
A possible explanation is that the constraint is more important for long sequence tasks than short sequence design since the search space is large and many mutations can easily go out of distribution.

\noindent \textbf{Adaptive-$\gamma$.}
BIB learns $\gamma$ and this leads to improvements over bidirectional mappings~($\gamma=0.5$) for all
tasks, verifying the effectiveness of Adaptive-$\gamma$.
We also consider the following variant,
%\vspace{-10pt}
\begin{equation}
    \begin{aligned}
    \boldsymbol{X}^{*} = \arg\min_{\boldsymbol{X}_{h}}
    \mathcal{L}_{bi}(\boldsymbol{X}_h, 0.5) - f_{aux}(\boldsymbol{X}_h)
    \,,
    \label{eq: joint}
    \end{aligned}
\end{equation}
which jointly optimizes the bidirectional learning loss $\mathcal{L}_{bi}(\boldsymbol{X}_h, 0.5)$ and the auxiliary term $f_{aux}(\boldsymbol{X}_h)$.
We found this yields similar or even worse results than pure bidirectional learning.
The reason may be that the weak supervision signal from $f_{aux}(\boldsymbol{X}_h)$ can serve as a guide to update the scalar $\gamma$ but not as a component of the main optimization objective that directly updates the sequence.

In the final column of Table~\ref{tab: ablation}, we examine the Adaptive-$\eta$ module.
Adding this module leads to improvements on all five tasks, which demonstrates its effectiveness.

\subsection{Adaptive-\texorpdfstring{$\eta$}{Lg}}
In this subsection, we aim to further demonstrate the effectiveness of the Adaptive-$\eta$ module on all six gradient-based methods.
We conduct experiments on two tasks: TFBind8(r) and avGFP.
Since the use of the infinitely wide neural network leads to poor
performance for BDI, we modify its implementation via deep linearization so that 
it can make use of the pre-trained LM. 

As shown in Table~\ref{tab: lr_learn}, \textit{Adaptive}-$\eta$ provides a consistent gain for all scenarios, which
demonstrates the widespread applicability and effectiveness of the module.
Furthermore, \textit{Adaptive}-$\eta$ leads to a maximum improvement of $1.4\%$ in
TFBind8(r) and $12.5\%$ in avGFP. ROMA is the algorithm that benefits
the most. 
One possible explanation is that ROMA incorporates a local smoothness
prior that leads to more stable gradients, with which \textit{Adaptive}-$\eta$ can be more effective.
Similar to Sec~\ref{subsec: abalation}, we consider the variant,
\begin{equation}
    \begin{aligned}
    \boldsymbol{X}^{*} = \arg\max_{\boldsymbol{X}_{h}}
    \mathcal{L}_{\boldsymbol{\theta}}(\boldsymbol{X}_h) + f_{aux}(\boldsymbol{X}_h)
    \,,
    \label{eq: joint1}
    \end{aligned}
\end{equation}
which performs joint optimization instead of bi-level optimization on two objectives.
As shown in Table~\ref{tab: lr_learn}, joint optimization generally deteriorates the performance. 
This again verifies that the auxiliary model can only serve as a guide instead of contributing to the main objective.

\begin{table*}
	\centering
	\caption{Adaptive-$\eta$ on all gradient-based methods.}  
	\label{tab: lr_learn}
	\scalebox{0.75}{
	\begin{tabular}{ccccccccccccc}
		\specialrule{\cmidrulewidth}{0pt}{0pt}
		\specialrule{\cmidrulewidth}{0pt}{0pt}
		\multirow{2}*{Method} & \multicolumn{6}{c}{TFBind8(r)} &  \multicolumn{6}{c}{avGFP} \\
		\cmidrule(lr){2-7}
        \cmidrule(lr){8-13}
		%\specialrule{\cmidrulewidth}{0pt}{0pt}
% 		\specialrule{\cmidrulewidth}{0pt}{0pt}
% 		\specialrule{\cmidrulewidth}{0pt}{0pt}
		 & \multicolumn{1}{c}{Grad} & \multicolumn{1}{c}{COMs}  & \multicolumn{1}{c}{ROMA}  & \multicolumn{1}{c}{NEMO}  & \multicolumn{1}{c}{BDI}  & \multicolumn{1}{c}{BIB} & \multicolumn{1}{c}{Grad} & \multicolumn{1}{c}{COMs}  & \multicolumn{1}{c}{ROMA}  & \multicolumn{1}{c}{NEMO}  & \multicolumn{1}{c}{BDI}  & \multicolumn{1}{c}{BIB} \\
% 		\cmidrule(lr){1-1}
% 		\cmidrule(lr){2-7}
%         \cmidrule(lr){8-13}
		\specialrule{\cmidrulewidth}{0pt}{0pt}
		Normal & ${0.941} $ & $0.921$ & ${0.926} $& $0.930$ &  $0.947$&  $0.952$ & $0.913$ &$0.938$  &  $0.975$ &  $0.914$ & $1.018$ & $1.060$ \\
% 		\cmidrule(lr){1-1}
% 		\cmidrule(lr){2-7}
%         \cmidrule(lr){8-13
        \specialrule{\cmidrulewidth}{0pt}{0pt}
		Joint & $0.941$ &$0.921$  &  $0.931$ &  $0.932$ & $0.935$ & $0.925$& $0.913$ &$0.905$  &  $0.923$ &  $0.906$ & $1.006$ & $1.009$ \\
		Gain & $0.000$ &$0.000$  &  $0.005$ &  $0.002$ & $-0.008$ & $-0.027$& $0.000$ &$-0.033$  &  $-0.052$ &  $-0.008$ & $-0.012$ & $-0.051$ \\
		\specialrule{\cmidrulewidth}{0pt}{0pt}
		Ada-$\eta$ & ${0.941} $ & $0.928$ & ${0.939} $& $0.935$ &  $0.951$& $0.956$& $0.916$ &$0.952$  &  $0.998$ &  $0.920$ & $1.024$ & $1.082$ \\
		Gain & $0.000$ &$0.007$  &  $0.013$ &  $0.005$ & $0.004$ & $0.004$& $0.003$ &$0.014$  &  $0.023$ &  $0.006$ & $0.006$ & $0.022$ \\
% 		\cmidrule(lr){1-1}
% 		\cmidrule(lr){2-7}
%         \cmidrule(lr){8-13}
        \specialrule{\cmidrulewidth}{0pt}{0pt}
		\specialrule{\cmidrulewidth}{0pt}{0pt}
	\end{tabular}
	}
\end{table*}
%\Sam{how to make this table more beautiful}

\section{Related Work}
\label{sec: related}

\begin{table*}
	\centering
	\caption{{Hyperparameter optimization from a gradient-based bi-level optimization view.}}  
	\label{tab: bi-level opt}
	\scalebox{.75}{
	\begin{tabular}{ccccc}
		\specialrule{\cmidrulewidth}{0pt}{0pt}
		Scenario & Inner variables & Inner Objective & Outer variables & Outer Objective \\
		\specialrule{\cmidrulewidth}{0pt}{0pt}
		model training & model params & minimize training loss~(e.g. cross-entropy loss) & hyperparams & minimize validation loss\\
        \specialrule{\cmidrulewidth}{0pt}{0pt}
            \specialrule{\cmidrulewidth}{0pt}{0pt}
		design optimization & design params & minimize training loss~(e.g. bidirectional learning  loss) & hyperparams & maximize auxiliary score \\	
		\specialrule{\cmidrulewidth}{0pt}{0pt}
	\end{tabular}
	}
\end{table*}
\noindent \textbf{Biological sequence design.}
There has been a wide range of algorithms for biological sequence design.
Evolutionary algorithms~\citep{sinai2020adalead, ren2022proximal} leverage the learned surrogate model to provide evolution guidance towards the high-scoring region.
\cite{angermueller2019model} propose a flexible reinforcement learning framework where sequence design is a sequential decision-making problem.
%, and use the learned surrogate as the reward during agent training.
%
Bayesian optimization methods propose candidate solutions via an acquisition function~\citep{terayama2021black}.
Deep generative model methods design sequences in the latent space~\citep{chan2021deep} or gradually adapt the distribution towards the high-scoring region~\citep{brookes2019conditioning}.
GFlowNets~\citep{jain2022biological} amortize the cost of search over learning and encourage diversity.
Gradient-based methods leverage a surrogate model and its gradient information to maximize the desired property~\citep{can2022bidirectional, norn2021protein,tischer2020design,linder2020fast}.
Our proposed BIB belongs to the last category and leverages the rich biophysical information
%from a pre-trained LM
~\citep{ji2021dnabert, elnaggar2020prottrans} to directly optimize the biological sequence.

\noindent \textbf{Offline model-based optimization.}
A majority of sequence design algorithms~\citep{angermueller2019model,sinai2020adalead, ren2022proximal} focus on the online setting where wet-lab experimental results in the current round are analyzed to propose candidates in the next round.
The problem of this setting is that wet-lab experiments are often very expensive, and thus a pure data-driven, offline approach is attractive and has received substantial research attention recently~\citep{trabucco2022design, kolli2022datadriven, beckham2022towards}. Gradient-based methods have proven to be effective~\citep{trabucco2021conservative, yu2021roma, fu2021offline, can2022bidirectional}.
Among these algorithms, \cite{can2022bidirectional} propose bidirectional mappings to distill information from the static dataset into a high-scoring design, which achieves state-of-the-art performances on a variety of tasks.
However, this bidirectional learning is designed for general tasks, like robot and material design,  and the rich biophysical information in millions of biological sequences is ignored.
In this paper, we leverage recent advances in deep linearization
to incorporate the rich biophysical information into bidirectional learning.
%from the pre-trained biological LM~\citep{ji2021dnabert, elnaggar2020prottrans} in bidirectional learning.

\textcolor{black}{
\noindent \textbf{Bi-level optimization for hyperparameter optimization.}
Gradient-based bi-level optimization~\citep{liu2021investigating, chen2022gradient} has been widely used in hyperparameter optimization to improve model training~\cite{maclaurin2015gradient, franceschi2017forward, pedregosa2016hyperparameter, lorraine2020optimizing, donini2019marthe, franceschi2018bilevel, chen2021generalized, luketina2016scalable, vicol2022implicit, micaelli2021gradient, bohdal2021evograd, baydin2017online}.
As shown in the \textit{model training} scenario of Table~\ref{tab: bi-level opt}, the inner level optimizes  model parameters by minimizing the training loss, and the outer level optimizes hyperparameters by minimizing the validation loss.
More specifically, \citep{maclaurin2015gradient} exactly reverse the optimization dynamics to compute hyperparameter
gradients.
\cite{luketina2016scalable} locally adjust
hyperparameters to minimize validation loss.
\citep{franceschi2018bilevel} unify hyperparameter optimization and meta-learning via bi-level optimization.
\citep{donini2019marthe} use past optimization information to simulate future behavior and compute the hypergradients efficiently.
All of these previous works aim to improve model training and belong to the \textit{model training} scenario in Table~\ref{tab: bi-level opt}. By contrast, our Adaptive-$\gamma$ and Adaptive-$\eta$ belong to the second scenario: \textit{design optimization}.}

\textcolor{black}{In our setting of offline model-based optimization, there is no validation set to provide hypergradient information to update hyperparameters, including the trade-off parameter $\gamma$ and the learning rate $\eta$.
Instead, inspired by previous work~\cite{angermueller2019model, trabucco2021conservative, chan2021deep} that uses an auxiliary model to select candidates, we use the auxiliary to provide a weak supervision signal for hyperparameter optimization.
This leads to a bi-level optimization task: the inner level optimizes design parameters by minimizing the training loss~(bidirectional learning loss), and the outer level optimizes hyperparameters by maximizing the auxiliary score.
As we can see, our formulation is different from that of previous work and thus can not compare with them directly.
}

% \noindent \textbf{Controlled text generation.}
% %
% A related field of work is controlled text generation, which aims to generate a natural text sequence with a predefined attribute.
% %
% PPLM~\citep{Dathathri2020Plug} combines a pre-trained LM with attribute classifiers to guide sequence generation by changing the hidden states.
% %
% Gedi~\citep{krause2020gedi} guides the generation of large LMs at each step by leveraging the supervision signals from the small LMs.
% %
% As pointed out in~\citep{chan2021deep}, the above methods cannot be directly applied to our task setting since they require the attribute to be within the training distribution.
% %
% GENH~\citep{chan2021deep} utilizes a regularized encoder-decoder framework and manipulates the learned latent space to generate desired sequences.

% \noindent \textbf{Bi-level optimization for hyperparameter tuning.}
% %
% Bi-level optimization has been widely used in hyperparameter tuning including learning rate~\citep{franceschi2017forward}, regularization~\citep{franceschi2018bilevel}, etc.
% %
% In this paper, we study tuning hyperparameters in offline model-based optimization, which, to the best of our knowledge, has not been investigated in previous work.
% %
% We leverage the weak supervision signal from an auxiliary model
% to update the hyperparameters via bi-level optimization.
% %
%\vspace{-10pt}

\section{Conclusion}
In this paper, we
propose bidirectional learning for offline model-based biological sequence.
Our work is built on the recently proposed bidirectional learning approach~\citep{can2022bidirectional}, which is designed for general inputs and relies on the NTK of an infinitely wide network to yield a closed-form loss computation.
Though effective, the NTK cannot learn features.
%because of the standard parametrization, and the adoption of NTK prevents the incorporation of powerful pre-trained LMs that can capture the rich biophysical information in millions of biological sequences. 
%
%Thus, 
We build a proxy model using the pre-trained LM model with a linear head and apply the deep linearization scheme to the proxy, which can yield a closed-form loss and incorporate the wealth of biophysical information at the same time.
In addition, we propose \textit{Adaptive}-$\gamma$ to  maintain a proper balance
between the forward mapping and the backward mapping by leveraging a weak supervision signal from an auxiliary model.
Based on this framework, we further propose \textit{Adaptive}-$\eta$, the first learning rate adaptation strategy compatible with all gradient-based offline model-based algorithms.
Experimental results on DNA and protein sequence design tasks verify the effectiveness of BIB and \textit{Adaptive}-$\eta$.

\textcolor{black}{We discuss potential negative impacts in Appendix~\ref{appendix: neg_impact}, reproducibility in Appendix~\ref{appendix: repro} and limitations of our research work in Appendix~\ref{appendix: limitation}.}

%For the DNA/protein benchmarks, we describe details in.

\normalem
\bibliography{example_paper}
\bibliographystyle{icml2023}

\section{Appendix}

\begin{table*}
	\centering
	\caption{Dataset details.}  
	\label{tab: details}
	\scalebox{0.93}{
	\begin{tabular}{ccccccc}
		\specialrule{\cmidrulewidth}{0pt}{0pt}
		\specialrule{\cmidrulewidth}{0pt}{0pt}
		Task &   Metric &   Min of $\mathcal{D}$ & Max of $\mathcal{D}$ & Min of $\mathcal{D}_{entire}$  & Max of $\mathcal{D}_{entire}$  \\
		\specialrule{\cmidrulewidth}{0pt}{0pt}
		TFBind8(r) & binding activity & $0.000$ & ${0.242} $& $0.000$ &  $1.000$\\
		TFBind10(r) & binding activity & $-1.859$ & ${-0.869} $& $-1.859$ &  $2.129$\\
		avGFP & fluorescence level & $1.283$ & ${2.175} $& $1.283$ &  $4.123$\\
		AAV & viruses viability & $-11.176$ & ${-1.814} $& $-11.176$ &  $9.536$\\
		E4B & ubiquitination rate  & $-3.589$ & ${-0.770} $& $-3.589$ &  $8.998$\\
		\specialrule{\cmidrulewidth}{0pt}{0pt}
		\specialrule{\cmidrulewidth}{0pt}{0pt}
	\end{tabular}
	}
\end{table*}

\begin{table}
	\centering
	\caption{{Experimental results on different pre-trained LMs for comparison.}}  
	\label{tab: pre_trained}
	\scalebox{.75}{
	\begin{tabular}{cccc}
		\specialrule{\cmidrulewidth}{0pt}{0pt}
		Pre-trained LM & avGFP & AAV & E4B\\
		\specialrule{\cmidrulewidth}{0pt}{0pt}
		ProtAlbert & $0.907\pm {0.004}$ & $0.478\pm {0.004}$ & $0.552\pm {0.023}$ \\
		ProtBert$_{\mathrm{(adopted)}}$ & ${1.060} \pm {0.016}$& $0.501 \pm {0.007}$& ${1.255} \pm {0.029}$ \\
		ProtBert-BFD & $1.119\pm {0.116}$ & $0.549\pm {0.009}$ & $1.880\pm {0.054}$ \\
		
		\specialrule{\cmidrulewidth}{0pt}{0pt}
	\end{tabular}
	}
 \vspace{-10pt}
\end{table}

\begin{table*}
	\centering
	\caption{{Experimental results on different size datasets for comparison.}}  
	\label{tab: diff_size}
	\scalebox{.75}{
	\begin{tabular}{cccccc}
		\specialrule{\cmidrulewidth}{0pt}{0pt}
		Dataset size & 20 & 40 & 60& 80 & 100 \\
		\specialrule{\cmidrulewidth}{0pt}{0pt}
		TFBind8(r) & $0.849\pm {0.027}$ & $0.883\pm {0.036}$ & $0.890\pm {0.033}$& $0.911\pm {0.042}$ & $0.923\pm {0.049}$ \\
		TFBind10(r) & ${0.248} \pm {0.000}$& $0.596 \pm {0.035}$& ${0.602} \pm {0.023}$& $0.616\pm {0.024}$ & $0.632\pm {0.036}$ \\
		\specialrule{\cmidrulewidth}{0pt}{0pt}
	\end{tabular}
	}
\end{table*}

{\begin{table*}
	\centering
	\caption{{Mean squared prediction losses for comparison.}}  
	\label{tab: mse}
	\scalebox{.75}{
	\begin{tabular}{cccccc}
		\specialrule{\cmidrulewidth}{0pt}{0pt}
		Method & TFBind8(r) & TFBind10(r) & avGFP & AAV & E4B\\
		\specialrule{\cmidrulewidth}{0pt}{0pt}
		Finetuned NN & ${0.101} \pm {0.001}$ & $1.130\pm {0.041}$ & $0.323\pm {0.006}$& $5.148\pm {0.074}$& $0.683\pm {0.012}$ \\
		Linearized NN & ${0.107} \pm {0.000}$ & $1.618\pm {0.000}$ & $3.956\pm {0.000}$& $23.041\pm {0.000}$& $1.050\pm {0.000}$ \\
		NTK & ${0.111} \pm {0.000}$ & $1.840\pm {0.000}$ & $4.866\pm {0.000}$& $24.451\pm {0.000}$& $1.075\pm {0.000}$ \\
    	\specialrule{\cmidrulewidth}{0pt}{0pt}
	\end{tabular}
	}
\end{table*}}

\subsection{DNA Embedding}
\label{appendix: dna}
To incorporate richer contextual information,
the DNA LM ~\cite{ji2021dnabert} adopts the $k$-mer sequence representation, which is widely used in DNA sequence analysis.
For example, the sequence $ATGGCT$ has its $3$-mer representation as $\{ATG, TGG, GGC, GCT\}$.
In this paper, we adopt its $3$-mer representation and compute the probability of the $3$-mer token by multiplying the probabilities of the three individual bases.
The $3$-mer representation is then sent to the pre-trained DNA LM.

%Our approach is somewhat different from the DNABert model~\citep{ji2021dnabert}, which adopts a $k$-mer sequence representation to incorporate richer contextual information.

\subsection{Different Pretrained LMs}
\label{appendix: diff_prt_lm}
{As shown in Table~\ref{tab: pre_trained}, we have tested the ProtBERT, ProtAlbert, and ProtBert-BFD models and found that better-quality models generally work better. The publicly available pre-trained DNA models are limited and thus we only perform experiments on the protein tasks. ~\cite{elnaggar2020prottrans} demonstrate that the language model performances follow the ordering: ProtBert-BFD $>$ ProtBert $>$ ProtAlbert. We can see that the performance ranks over the three protein tasks avGFP, AAV, and E4B are the same in Table~\ref{tab: pre_trained}.}

\subsection{Task Details}
\label{appendix: dataset}

{We conduct experiments on two DNA tasks following~\citep{can2022bidirectional} and three protein tasks in \citep{ren2022proximal} which have the most data points. We report the dataset details in Table~\ref{tab: details}.}

{\noindent \textbf{DNA Task 1 TFBind8(r).}
The goal is to find a length-$8$ DNA sequence to maximize the binding activity score with a particular transcription factor,
$\rm{SIX6 REF R1}$~\citep{barrera2016survey}. %The ground-truth binding affinities of all possible sequences are available in.
%
%Each nucleotide is one of four categorical variables and 
We sample $5000$ data points for the offline algorithms following~\citep{can2022bidirectional}.}

{\noindent \textbf{DNA Task 2 TFBind10(r).}
The task TFBind10(r) is the same as TFBind8(r) except that the goal is to find a length-$10$ DNA sequence.
Both DNA tasks measure the entire search space and we adopt these measurements as the approximate ground-truth evaluation.}

{\noindent \textbf{Protein Task 1 avGFP.}  
This task aims to find a protein sequence with approximately 239 amino acids to maximize the fluorescence level of Green Fluorescent Proteins~\citep{sarkisyan2016local}. 
%
%Each entry in the protein sequence represents one amino acid and takes on one of the $20$ values.
%
The task oracle is constructed by using the full unobserved dataset (around 52,000 points) following~\citep{ren2022proximal}.
The oracle passes the average of the residue embeddings from the pre-trained Prot-T5 ~\citep{elnaggar2020prottrans} into a linear layer and then fits the dataset.
The following two task oracles take the same form.
Offline algorithms can only access the lowest-scoring 26,000 points.}

{\noindent \textbf{Protein Task 2 AAV.}  The goal is to engineer a $28$-amino
acid segment (positions $561$--$588$) of the VP1 protein to remain viable for gene therapy~\citep{bryant2021deep}.
We use the entire $284,000$ data points to build the oracle and the lowest-scoring $142,000$ points for the offline algorithms.}

{\noindent \textbf{Protein Task 3 E4B.}
This task aims to design a protein~(around 102 amino acids) to maximize the ubiquitination rate to the target protein~\citep{starita2013activity}.
The full dataset with around $100,000$ points is used to build the oracle and the bottom half is used for the offline algorithms.}

\cite{lee2023protfim} evaluate the sequence generation and structure conservation simultaneously with the help of Alphafold2, which provides a more accurate evaluation.

\noindent \textbf{Oracle Parameterization.}
{The parameterization of the oracle is different from that of the regression model from two aspects: 1) model architecture; 2) pre-trained information source. First, the oracle adopts the Prot-T5 model which consists of an encoder and a decoder, while the regression model adopts the Prot-BERT model which only has an encoder. Second, Prot-T5 is trained on the BFD and UniRef100 datasets and ProtBert is trained on the UniRef50 dataset. These two points demonstrate that the oracle and the regression model are different function classes. We choose the Prot-T5 model as the oracle because this is the state-of-the-art protein LM to extract features and recent work~\cite{elnaggar2020prottrans} has demonstrated its effectiveness. In order to test how related the Prot-T5 (oracle)/Prot-BERT(proxy) models are, we trained them on a sampled training dataset and compared the test predictions of the testing set. By evaluating the Pearson correlation coefficient (PCC) between the two prediction errors PCC(ProtT5 predictions - test labels, ProtBERT predictions -test labels), we obtain $0.0104$ on avGFP, $-0.0005$ on AAV, and $-0.0062$ on E4B. These results suggest that the two models are not strongly related in terms of the predictions they form. }

{Following~\citep{trabucco2021conservative}, we select the top ${N}=128$ most promising sequences for each comparison method.
Among these sequences, we report the maximum  normalized ground truth score as the evaluation metric following~\citep{ren2022proximal}.}

\subsection{Different Dataset Size}
\label{appendix: diff_data_size}
{As shown in Table~\ref{tab: diff_size}, we have tested the performance of BDI as a function of dataset size (N= $20$, $40$, $60$, $80$, $100$) in TFBind8(r) and TFBind10(r) since they have exact oracle evaluations. We see that performance is already good for N=$20$ for TFBind8(r) and N=$40$ for TFBind10(r).}

\subsection{Training Details}
\label{appendix: init}
We use Pytorch~\citep{paszke2019pytorch} to run all experiments on one V100 GPU.
Following the setting in~\cite{norn2021protein}, we introduce a length-$L$ protein sequence as a continuous random matrix $\boldsymbol{X}_h \in R^{L\times 20}$~($\boldsymbol{X}_h \in R^{L\times 4}$ for DNA), initialized using a normal distribution with the mean $0$ and the standard deviation of $0.01$.
To make this sequence correspond correctly to the candidate, we exchange the largest value in $\boldsymbol{X}[l, :]$ with the value in the amino acid index.

\subsection{Ranking Performance}
\label{appendix: rank_perf}
{As for prediction performances, the rank should be: a NN $>$ linearized pre-trained LM $>$ NTK. We have conducted experiments to verify this. We sample half of the data, train a model to predict another half data, and report the mean squared loss here and in Table~\ref{tab: mse}. A small mean squared loss indicates a good prediction performance; thus, we have verified the above ranking order.}

\subsection{Negative Impact}
\label{appendix: neg_impact}
Protein sequence design aims to find a protein sequence with a particular biological function, which has a broad application scope.
This can lead to improved drugs that are highly beneficial to society.
For instance, designing the antibody protein for SARS-COV-2 can potentially save millions of human lives~\citep{kumar2021drug} and designing novel anti-microbial peptides~(short protein sequences) is central to tackling the growing public health risks caused by anti-microbial resistance~\citep{murray2022global}.
Unfortunately, it is possible to direct the research results towards harmful purposes such as the design of biochemical weapons.  
As researchers, we believe that we must be aware of the potential harm of any research outcomes, and carefully consider whether the possible benefits outweigh the risks of harmful consequences. 
We also must recognize that we cannot control how the research may be used.
In the case of this paper, we are confident that there is a much greater chance that the research outcomes will have a beneficial effect. 
We do not consider that there are any immediate ethical concerns with the research endeavour. 

\subsection{Reproducibility Statement}
\label{appendix: repro}
We provide the code implementation of BIB and \textit{Adaptive}-$\eta$ ~\href{https://anonymous.4open.science/r/BIB-ICML2023-Submission/README.md}{here} and we also attach the code in the supplementary material.
We describe DNA/protein benchmarks in Sec.~\ref{subsec: benchmark} and training details in Sec.~\ref{subsec: training_details}.
We explain how to obtain the sequence embedding from the pre-trained LM and how to perform gradient ascent on the sequence in Sec.~\ref{sec: pre}.

\subsection{Limitations}
\label{appendix: limitation}
\textcolor{black}{We note that, due to the nature of offline optimization, we propose designs that are outside the available datasets. Evaluation of the performance of such designs must then rely on training an oracle model based on the entire available data. We are careful to ensure that (i) the oracle is significantly structurally different from the model(s) used within the optimization algorithm, (ii) the oracle is accurate, especially in terms of rank correlation; and (iii) oracle residuals and regression model residuals are uncorrelated. However, even after these measures are taken, there is a danger that the proposed designs are not biologically optimal, i.e., that the oracle performance is not genuinely reflective of performance that would be observed in experiments. This concern can only be alleviated via further biological experiments that examine how well off-line optimization algorithms perform in practice.}

\end{document}